\documentclass[twocolumn,nofootinbib,preprintnumbers,amsmath,amssymb]{revtex4}

\usepackage{graphicx}
\usepackage{dcolumn}
\usepackage{bm}
\usepackage{amsmath}

\def \be {\begin{equation}}
\def \ee {\end{equation}}

\begin{document}

\title{Pulsars as Standard Candles}

\author{Andrei Gruzinov}

\affiliation{ CCPP, Physics Department, New York University, 4 Washington Place, New York, NY 10003
}

\begin{abstract}

To illustrate the standard candle property of gamma-ray pulsars (and also to thereby confirm the recent first-principle calculation of pulsar gamma-ray emission), we ``measure'', via the lightcurve fitting, three distances and one moment of inertia of some weak pulsars. We are not sure what the three distances are good for, but the measurement of the moment of inertia must be of interest for nuclear physics. 

Although we must state that the quality of the numerical program which we use to calculate the lightcurves and efficiencies is inadequate (as are the author's qualifications as a numericist and data analyst), in good hands, and upon an easily doable extension to non-weak pulsars, the method's yield should be impressive.

\end{abstract}

\maketitle

\section{Introduction}

For three (supposedly weak) pulsars from the Fermi Pulsar Catalog \cite{Fermi2013}, PSRs J1057-5226, J1741-2054, J2055+2539, we measure\footnote{For reasons stated in the Abstract, the following numbers have the status of an illustration.}, to some 10\% accuracy, the distances:
\be
d=0.41,~0.37,~0.35I_{45}^{1/2}{\rm kpc},
\ee
respectively; $I_{45}$ are the moments of inertia in units of $10^{45}{\rm g}\cdot{\rm cm}^2$. The Catalog lists these distances as 
\be
d=0.3\pm0.2,~0.38\pm0.02,~<15.3{\rm kpc},
\ee
respectively. Then, for PSR J1741-2054, we have
\be
I_{45}=1.1\pm0.2~~~.
\ee

Have we had better numerics, we would have measured absolute distances (and moments of inertia, \cite{Gruzinov2013}(c)) directly from Fermi and checked the observers' suspiciously accurate distance to PSR J1741-2054. (The theory of gamma-ray emission turned out to be so simple, that, after good coding, it should become more reliable than the dispersion-measure distance determination.)

As it is, due to poor numerical accuracy (Appendix), we do the measurement by a downgraded procedure which uses only lightcurves and efficiencies, but not the spectra; this gives only the ratio $d/I^{1/2}$. 

The measurement procedure is outlined in \S\ref{procedure}; the implementation and results are presented in \S\ref{results}.

\section{Procedure}\label{procedure}

From the Catalog we take the following:
\begin{itemize}

\item period $P$;
\item period derivative $\dot{P}$;
\item bolometric flux $f$ ($[f]={{\rm erg}\over {\rm cm}^2{\rm s}})$;
\item bolometric lightcurve $l_{\rm obs}(\phi )$, where $\phi$ is the pulse phase; $l_{\rm obs}(\phi )$ is proportional to the bolometric flux at a given phase, and normalized by $l_{\rm obs~max}=1$. Catalog lightcurves are actually in ``W. Counts/bin'', and we are not sure how to translate this into a bolometric lightcurve; but as our procedure  uses only some seemingly robust morphological characteristics of the lightcurves, we do not expect a large error from this uncertainty. (The chosen pulsars have lightcurves with pronounced morphology).

\end{itemize}

\begin{figure}[bth]
  \centering
  \includegraphics[width=0.48\textwidth]{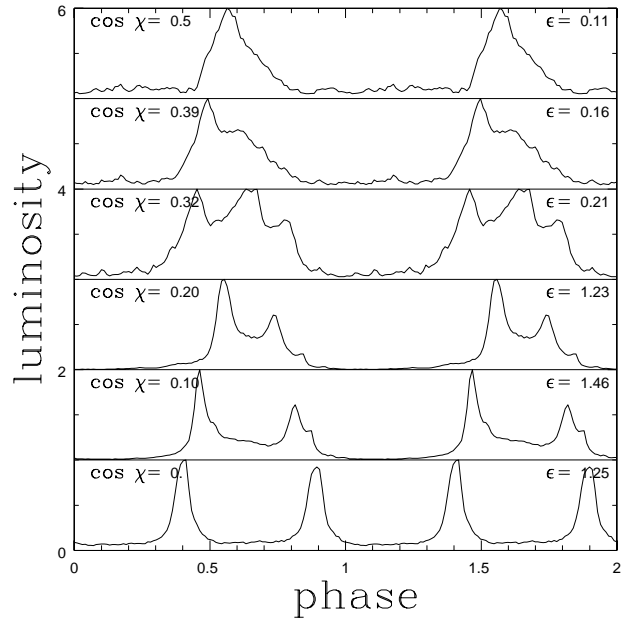}
\caption{Lightcurves of a weak pulsar; spin-dipole angle $\theta =26^{\circ }$ ($\cos \theta =0.9$). The labels indicate the observation angle and the efficiency $\epsilon (\chi , \theta )$. Following observers' convention, we plot two periods, and normalize the full phase to one. } \label{curves1}
\end{figure}

For weak pulsars, the 'theory' gives the following:
\begin{itemize}
\item The normalized bolometric lightcurve is 
\be\label{curve} 
l_{\rm th}(\phi )=l(\phi ; \chi, \theta ).
\ee
Here $\theta$ is the spin-dipole angle, $\chi$ is the observer angle (the angle between the spin axis and the direction to observer).
\item Bolometric efficiency
\be\label{eff} 
\epsilon =\epsilon(\chi, \theta),
\ee
defined as the ratio of the pulsed bolometric luminosity (as seen at observation angle $\chi$) to the spin-down power. 

\end{itemize}

The measurement procedure is then: 

\begin{itemize}

\item Use Eq.(\ref{curve}) to fit the lightcurve, thereby measuring both $\theta$ and $\chi$.
\item Use Eq.(\ref{eff}) to calculate the bolometric luminosity (in terms of the known $P$ and $\dot{P}$ and unknown $I$), and then, knowing the bolometric flux $f$, deduce the distance to the pulsar, $d$.

\end{itemize}

\section{Results}\label{results}

We use the numerical procedure of \cite{Gruzinov2013}(b) to calculate theoretical bolometric lightcurves $l(\phi ; \chi, \theta )$ and efficiencies $\epsilon(\chi, \theta)$. We then perform a visual fit of the lightcurve morphology: for each calculated magnetosphere (of various spin-dipole angles $\theta$), we adjust an observation angle $\chi$ so as to make the lightcurve single-peak and as ``pedestal''-like as possible, or single-peak and as ``high-chair''-like as possible.

PSRs J1057-5226 and J1741-2054, at our level of accuracy, are both single-peak ``pedestals'', although of a different full width at half-maximum, $\delta \phi=0.34$ and 0.29, respectively; PSR J2055+2539 is a single-peak ``high-chair'' of width  $\delta \phi=0.26$.

We get satisfactory morphology fits at various spin-dipole angles; but as shown in the Table, these best fits come with different full widths at half-maximum $\delta \phi$.

\begin{table}[bth]
\centering
\begin{tabular}{ccccccccc}

$\cos \theta$ &~~~& $\cos \chi$ & $\epsilon$ & $\delta \phi$ &~~~~~~ &$\cos \chi$ & $\epsilon$ & $\delta \phi$ \\

\\

0.9 &~~~& 0.32 & 0.21 & 0.41&~~~~~~& 0.39 & 0.16 & 0.26 \\
0.8 &~~~& 0.41 & 0.19 & 0.25&~~~~~~& 0.46 & 0.13 & 0.23 \\
0.7 &~~~& 0.50 & 0.14 & 0.22&~~~~~~& 0.53 & 0.11 & 0.19 \\

\end{tabular}
\caption{\label{pedestal} Best fits of ``pedestal'' (left) and ``high-chair'' (right) morphologies. For each spin-dipole angle $\theta$, listed are the best-fitting observation angle $\chi$, the corresponding efficiency $\epsilon$, and the full width at half-maximum $\delta \phi$.  }
\end{table}

For each pulsar, a unique (interpolated) pair $(\chi , \theta )$ is selected from the Table by fitting $\delta \phi$. Then we use the (interpolated from the Table) efficiency, $\epsilon$, to calculate the bolometric luminosity and hence the distance:

\begin{tabular}{cccc}
\\
PSR & J1057-5226 & J1741-2054 & J2055+2539 \\
\\
$P$ (ms) & 197 & 414 & 320  \\
$\dot{P}$  ($10^{-15}$) & 5.8 & 17.0 & 4.1 \\
$L_{\rm sd}$ ($10^{34}I_{45}{{\rm erg}\over {\rm s}}$) & 3.0  & 0.94  & 0.49 \\
$\epsilon$ & 0.20 & 0.20 & 0.16 \\
$L_{\rm bol}$ ($10^{33}I_{45}{{\rm erg}\over {\rm s}}$) & 6.0 & 1.9 & 0.79 \\
$f_{\rm bol}$ ($10^{-11}{{\rm erg}\over {\rm cm}^2{\rm s}}$) & $29.5\pm0.3$ & $11.7\pm0.4$ & $5.4\pm0.2$ \\
\\
$d$ ($I_{45}^{1/2}$kpc) & 0.41 & 0.37 & 0.35\\
\\

\end{tabular}

We place an accuracy on our results in the following (arbitrary) manner:
\begin{itemize}
\item In the worst of the calculated cases ($\cos \theta =0.9$), the Poynting flux through the sphere of radius 0.5 (light cylinders), $L_{\rm in}$, the Poynting flux through the sphere of radius 2.3, $L_{\rm out}$, and the radiation power emitted from the corresponding spherical shell, $L_{\rm damp}$, were mismatched by 16\% , with $L_{\rm out}/L_{\rm in}=0.49$, and $L_{\rm damp}/L_{\rm in}=0.35$. An approximately 20\% error in efficiency gives an approximately 10\% error in the distance.
\item At each spin-dipole angle $\theta$, a  satisfactory fit of morphology exists in an interval of observation angles $\delta \chi \sim 0.05$, and within this interval $\epsilon (\chi ,\theta)$ changes by about 10\%.
\item The box ($6^3$ big in our case) is too small and must contaminate our results. The star is poorly resolved, and emission from the near star region must be excluded by hand. The lightcurves shown in the figures were computed counting only the photons emitted in the spherical shell described above. Moving the walls of this arbitrarily selected spherical shell changes our results by about 10\%.\footnote{ The radii of the emitting spherical shell, 0.5 and 2.3, were (arbitrarily) chosen as follows. From the axisymmetric calculation, we know that not much is emitted within 0.5, but there is some emission all the way to about 3. We however have a box at 3, we need to step away from it. The outer radius 2.3 gives an about correct mean efficiency for $\theta =0$.}
\end{itemize}

\begin{figure}[h]
  \centering
  \includegraphics[width=0.48\textwidth]{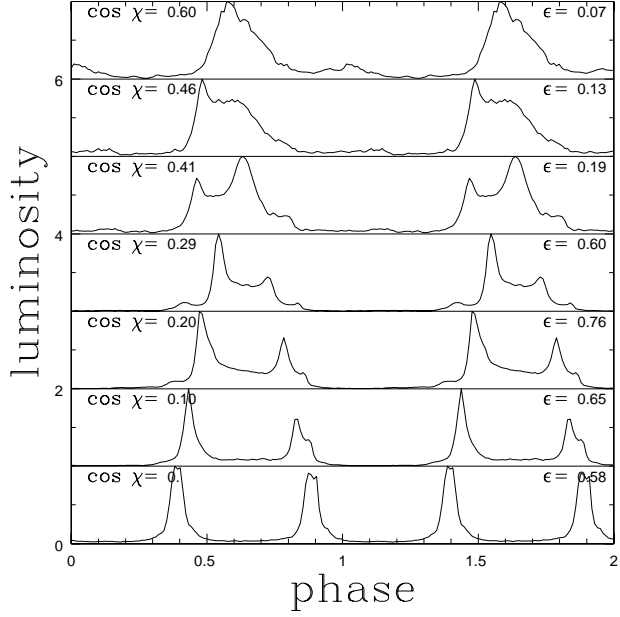}
\caption{Spin-dipole angle $\theta =37^{\circ }$ ($\cos \theta =0.8$).} \label{curves2}
\end{figure}

\section{Conclusions}

\begin{itemize}

\item The Aristotelian Electrodynamics (first-principle) calculation of pulsars passes all tests this author is capable of reliably performing.
\item Better numerics should allow more tests (and maybe uses) of the theory as applied to weak pulsars.
\item The theory should be extended to non-weak pulsars; this appears easily doable. 
\end{itemize}

\begin{acknowledgments}

I thank John Kirk for pointing out that the basic equation of Aristotelian Electrodynamics was written (in a different form and with a different derivation) in \cite{Finkbeiner1989}.

\end{acknowledgments}

\begin{figure}[bth]
  \centering
  \includegraphics[width=0.48\textwidth]{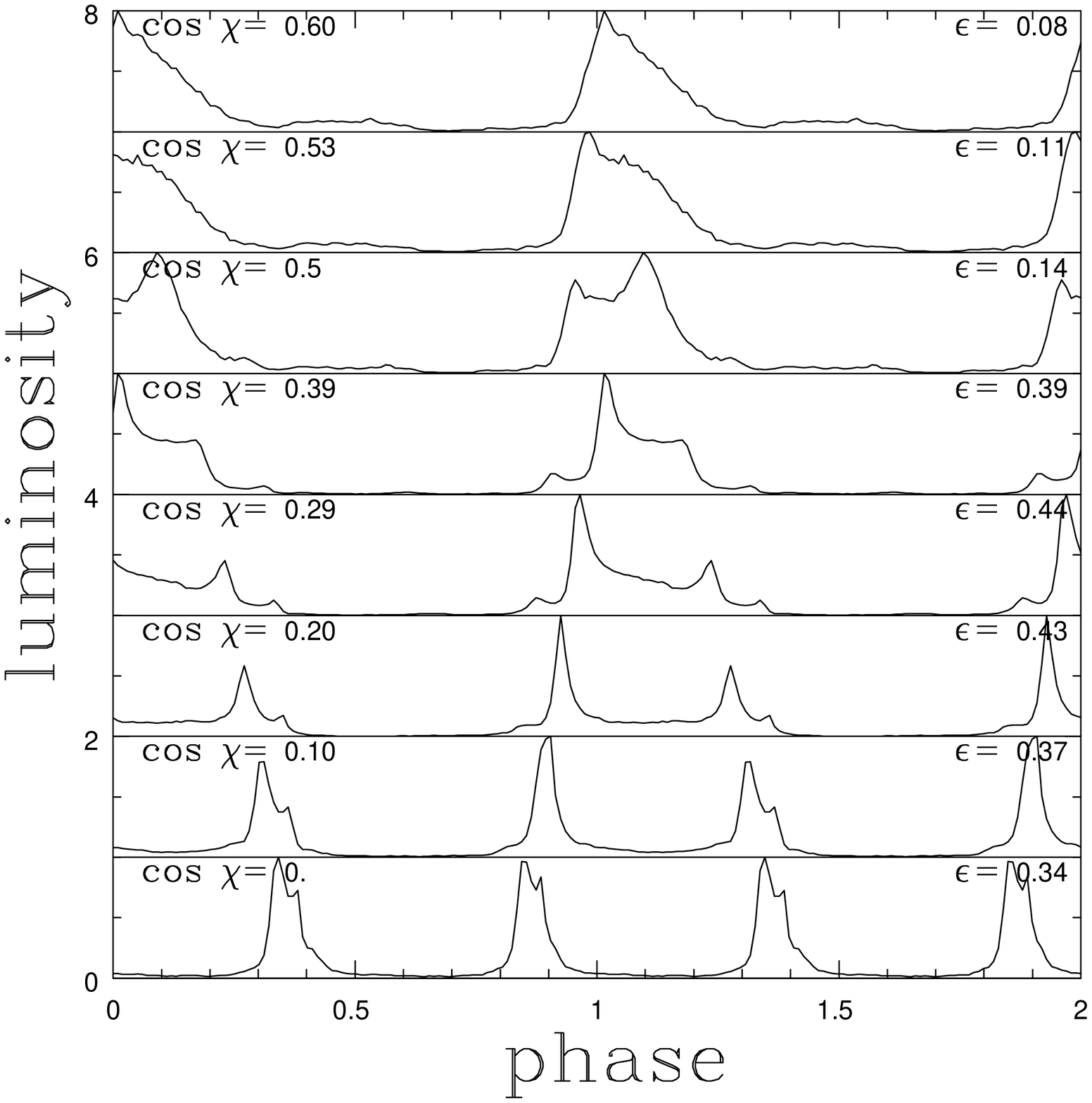}
\caption{Spin-dipole angle $\theta =46^{\circ }$ ($\cos \theta =0.7$).} \label{curves3}
\end{figure}

\begin{figure}[bth]
  \centering
  \includegraphics[width=0.48\textwidth]{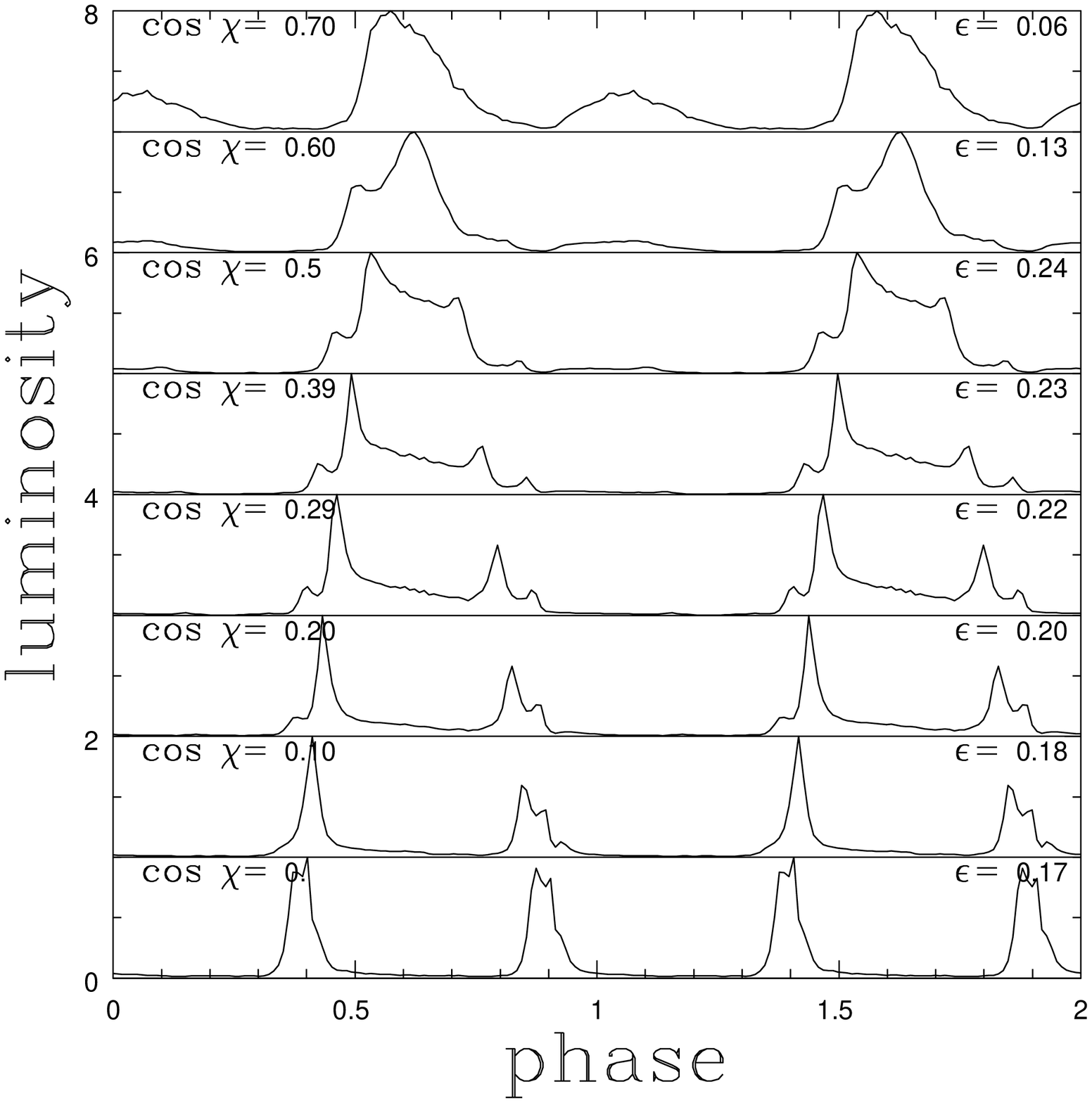}
\caption{Spin-dipole angle $\theta =60^{\circ }$ ($\cos \theta =0.5$).} \label{curves4}
\end{figure}

\begin{figure}[bth]
  \centering
  \includegraphics[width=0.48\textwidth]{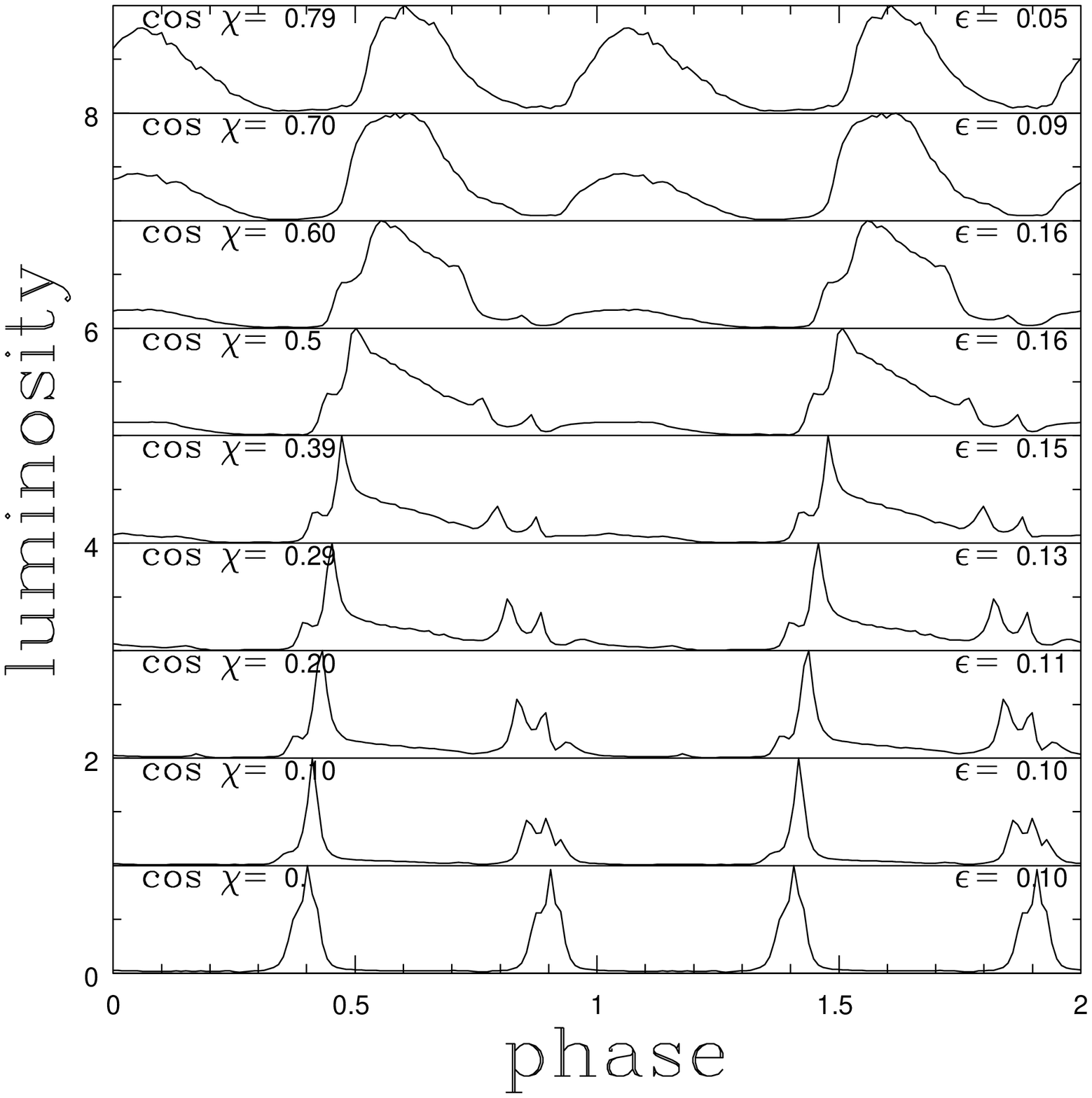}
\caption{Spin-dipole angle $\theta =73^{\circ }$ ($\cos \theta =0.3$).} \label{curves4}
\end{figure}

\begin{figure}[bth]
  \centering
  \includegraphics[width=0.48\textwidth]{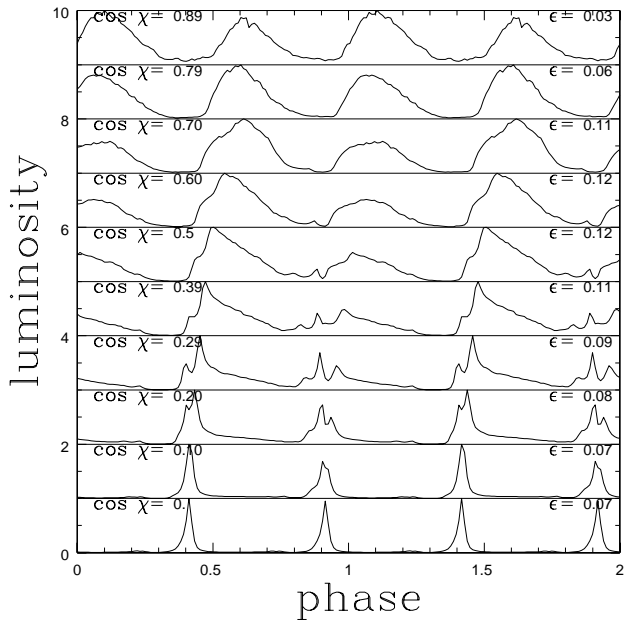}
\caption{Spin-dipole angle $\theta =84^{\circ }$ ($\cos \theta =0.1$).} \label{curves4}
\end{figure}

\appendix* 
\section{Why we were unable to calculate the spectrum in three dimensions}

The way we code, we get convergence of the calculated emission spectrum in axisymmetry only at 1600x3200 resolution \cite{Gruzinov2013}(a). We also find that at least a 5x10 (light cylinders) box is needed, and the star radius, $r_s$, should be smaller than about 0.2. 

In three dimensions \cite{Gruzinov2013}(b), we calculate an $r_s=0.33$ star, at a $270^3$ resolution, in a $6^3$ box, which corresponds to a 135x270 resolution in a 3x6 axisymmetric box. To gauge our results, we have calculated an aligned pulsar by the three-dimensional code; comparing the results to the high-resolution axisymmetric simulation, we see that the accuracy of the calculated curvature is poor, and hence we cannot reliably calculate the spectrum for the computed three-dimensional magnetospheres. 

On the other hand, the computed electromagnetic field (and hence velocities of charges, and hence the lightcurves) and also the calculated efficiencies are close in the two simulations. We have therefore decided to use only the calculated efficiencies and lightcurves, hoping that the photon cutoff energy (which was reliably calculated only in axisymmetry) is not very sensitive to the spin-dipole angle (at fixed spin-down power); this needs to be checked.


\begin{thebibliography}{99}

\bibitem{Fermi2013}
The Fermi-LAT collaboration, arXiv:1305.4385 (2013)

\bibitem{Gruzinov2013}
A. Gruzinov (a,b,c),  arXiv:1309.6974, 1310.1894, 1310.3261 (2013)

\bibitem{Finkbeiner1989}
B. Finkbeiner, H. Herold, T. Ertl, H. Ruder, Astron. Astrophys. {\bf 225}, 479 (1989)


\end{thebibliography}
\end{document}